\documentclass[12pt]{article}

\newcommand{\bea}{\begin{eqnarray}}
\newcommand{\eea}{\end{eqnarray}}
\newcommand{\pa}{\partial}
\newcommand{\bi}{{\bar{i}}}

\newcommand{\bk}{{\bar{k}}}

\newcommand{\bz}{{\bar{z}}}

\newcommand{\bV}{{\bar{V}}}
\newcommand{\fb}{{\bar{f}}}
\newcommand{\bZ}{{\bar{Z}}}
\newcommand{\bF}{{\bar{F}}}
\newcommand{\bmu}{{\bar{\mu}}}
\newcommand{\bd}{{\bar{d}}}
\def\a{\alpha}
\def\href#1#2{#2}
\begin{document}
\begin{titlepage}
\hfill
\vbox{
    \halign{#\hfil         \cr
           hep-th/0406165 \cr
           IHES/P/04/30 \cr
           } 
      }  
\vspace*{30mm}
\begin{center}
{\Large \bf Holomorphically Covariant Matrix Models}
\vspace*{15mm}

{FURUUCHI \ Kazuyuki} 
\vspace*{1cm}

{Department of Physics and Astronomy,
University of British Columbia,\\
6224 Agricultural Road, 
Vancouver, B.C., V6T 1Z1, Canada}\\
\vspace*{5mm}
{Institute des Hautes Etudes Scientifiques,\\ 
35 route de Chartres, 91440 Bures-sur-Yvette, France}
\vspace*{10mm}
\end{center}
\begin{abstract}
\noindent We present a method to construct
matrix models on arbitrary simply connected
oriented real two dimensional
Riemannian manifolds.
The actions and the path integral measure 
are invariant under holomorphic 
transformations of matrix coordinates.
\end{abstract}

\end{titlepage}

\vskip 1cm
\newpage

\section{Introduction}

One of the most fascinating aspects of
string theory is that it is
a quantum theory that contains gravity.
The discovery of Dirichlet branes \cite{D} has 
revealed the importance of the view-point that 
gravitational theories which are perturbatively
described by closed strings may have
dual descriptions in terms of open strings.
A peculiar feature of 
the open string description of D-branes is
that the positions of multiple D-branes 
are described by matrices
\cite{witten}.
When those matrices are
simultaneously diagonalizable,
The eigenvalues have a natural interpretation 
as the positions of the D-branes.
However, such an interpretation becomes obscure
when those coordinate matrices do not
commute.  This non-abelian nature of matrix coordinates
introduces non-locality
which is typically exhibited in the
formation of higher dimensional branes \cite{Ish1,Myers}.
This in turn may resolve problems
arising from classical singularities in general relativity.
Since open string degrees of freedom become
important at the sub-stringy scale \cite{short},
we expect it to capture fundamental structures of
space-time in the microscopic scale.
This led to matrix model proposals where
matrix coordinates are the fundamental variables \cite{BFSS,IKKT}.
On the other hand, it does not seem
straightforward to incorporate the 
non-locality with
local symmetries of the target space, in particular
diffeomorphisms, 
the classical symmetries of general relativity.
But at least at 
sufficiently weak string coupling and low energy,
diffeomorphisms should be realized in
actions describing multiple D-branes 
on general gravitational backgrounds.
Diffeomorphism symmetry will also be a key to understand
how curved space-time will be realized in matrix models.
\footnote{Our motivation for studying this issue 
was affected by communications with
researchers in type IIB matrix model.
See \cite{space,polymer,sugino,azuma} for other approaches
on this issue.
}

Construction
of multiple D-brane actions on
K\"ahler manifolds
was studied earlier based 
on Douglas' geodesic distance criterion,
a requirement that
the mass of fluctuations around a diagonal configuration be  
equal to the shortest geodesic distance 
between the diagonal elements
\cite{douglas,douglas2,douglas3}.
Recently, approaches based on
covariance under 
general coordinate transformations have been developed
\cite{dbs,dbsw,ours},\footnote{See also \cite{blending} 
for a case of local gauge symmetry in flat space.
Some ideas and techniques appeared 
there were precursors of 
those in \cite{ours} and this article.} 
where actions that satisfy the geodesic distance criterion 
should in principle 
be obtained as special cases.
However, in these works
the complete form of the matrix coordinate transformation,
or even its existence in a rigorous sense, were not shown.  
This remains as a challenging issue.
Under these circumstances, it will be
important to show that
there exists a simple well-defined quantum model
where the matrix coordinate transformation law
is completely specified.

In this article, we construct matrix models 
on arbitrary simply connected oriented 
real two dimensional
Riemannian manifolds.  
The generalization of
holomorphic coordinate transformations
to matrix coordinates is straightforward 
in this case.
It is still non-trivial 
to construct covariant actions which in general
contain 
holomorphic and anti-holomorphic matrix coordinates,
in particular they will contain commutators
of holomorphic and anti-holomorphic matrix coordinates.
Further, we construct an invariant
path integral measure 
which is used for defining
the quantum models.

\section{Holomorphically covariant matrix models}

\subsection{Holomorphic transformations
for matrix coordinates}

\subsubsection*{\it Holomorphic transformations}

Let us consider an
oriented 
real two dimensional Riemannian
manifold ${\cal M}$.
Note that such manifolds are always K\"ahler.
Suppose we have a K\"ahler potential 
$K(z,\bz)$ on ${\cal M}$,
where $z$ and $\bz$ are holomorphic and anti-holomorphic
coordinates, respectively.
In this article, we restrict ourselves
to the case where ${\cal M}$
is simply connected and we assume that
the coordinate $z$ covers the whole ${\cal M}$.
We'd like to promote the holomorphic coordinate 
to $N \times N$ complex
matrix coordinate $Z$.
In analogy with D-brane theories,
we call $N$ the number of D-instantons.
We may need to take $N$ to infinity in some
appropriate scaling limit
in order to define a matrix model relevant to nature,
but here we are interested in the mathematical 
structure of matrix coordinate transformations
which are common to multiple D-brane actions.
For a holomorphic coordinate transformation
\bea
\label{holo}
z \rightarrow f(z)
\eea
we define a corresponding holomorphic 
matrix coordinate transformation from 
$Z$ to $F$
through the Taylor series
\bea
\label{holom}
Z \rightarrow 
F(Z) = 
f(Z) \equiv
\sum_{n=0}^\infty \frac{1}{n!} \pa_z^n f(z_0) (Z-z_0)^n .
\eea
For the anti-holomorphic coordinate $\bz$,
we associate a matrix $\bZ$
which is the hermitian conjugate matrix of $Z$.
We restrict ourselves 
to holomorphic transformations
for the holomorphic coordinate, and 
anti-holomorphic transformations
for the anti-holomorphic coordinate.
The transformation involves
only one matrix coordinate so there is no matrix
ordering ambiguity.
Requiring that when the coordinate
matrix is diagonal
the transformation reduces to the usual coordinate transformation 
for each diagonal element, 
the simple rule (\ref{holom}) is the only possibility.
This simplifies the problem drastically
compared to the higher dimensional case.
As a nature of Taylor series, the definition (\ref{holom})
does not depend on the expansion point $z_0$.
It also satisfies the
composition law
\bea
 \label{composition}
H(F(Z)) = 
H \circ F (Z) \equiv
h \circ f (Z) \, .
\eea
In one complex dimension
both independence of the expansion point and
the composition law are realized simply, whereas
in higher dimensions it is a challenging issue to 
incorporate both of them simultaneously.

\subsubsection*{\it Defined region}

When the coordinate $z$ is defined
in a region
$\Omega_z$, 
we define defined region of the corresponding 
matrix coordinate $Z$
by requiring its eigenvalues 
to be in the region $\Omega_z$:
\bea
 \label{spec}
 spec\, Z \subset \Omega_z \, .
\eea
Suppose the region
is mapped as
$\Omega_z \rightarrow \Omega_f$
under a holomorphic map $z \rightarrow f(z)$.
If $Z$ satisfies (\ref{spec}), then $F$ satisfies
\bea
 \label{specF}
 spec\, F \subset \Omega_f \, .
\eea
Therefore, the definition of the
defined region of matrix coordinate
is consistent with holomorphic coordinate
transformations.
The definition (\ref{spec}) is also
invariant under
$U(N)$ transformation 
\bea
 \label{UN}
 Z \rightarrow UZU^{\dagger}, \quad  
\bZ \rightarrow U\bZ U^{\dagger}
\eea
where $U$ is an $N\times N$ unitary matrix:
\bea
 \label{specU}
 spec\,\, UZU^\dagger = spec\, Z \,.
\eea
We will loosely denote $spec\, Z \subset {\cal M}$
when all the points whose positions in
coordinate $z$ are eigenvalues of $Z$
are on ${\cal M}$,
by identifying points with their coordinates.

\subsubsection*{\it K\"ahler normal coordinates}

In our construction of
generally covariant actions for multiple D-branes
in \cite{ours}, a
crucial role was played by a matrix-valued object 
which 
transforms as a tangent vector
under coordinate transformations.
It can be regarded as a matrix
generalization of Riemann normal coordinates.
Since here we are 
considering 
holomorphic coordinate transformations instead,
we need an object which
transforms as a holomorphic tangent vector.
Such a vector for ordinary K\"ahler manifolds
was constructed in \cite{HN,HIN}
and called K\"ahler normal coordinates.
K\"ahler normal coordinates differ from Riemann normal coordinates
by the anti-holomorphic part of the Riemann normal coordinates.
The K\"ahler normal
coordinate $v^z_{z_0}$ at point $z_0$
is defined via the holomorphic coordinate transformation
\bea
 v^z_{z_0}(z)
 &=& z-z_0 + \sum_{n=2}^{\infty} \frac{1}{n!}
 g^{z\bz}(z_0,\bz_0) \pa_z^n \pa_\bz K(z_0,\bz_0)
   (z-z_0)^n \nonumber \\
 &=&  \sum_{n=1}^{\infty} \frac{1}{n!}
 g^{z\bz}(z_0,\bz_0) \pa_z^n \pa_\bz K(z_0,\bz_0)
   (z-z_0)^n 
 \label{KNC}
\eea
where the K\"ahler metric is given by
$g_{z\bz}(z,\bz) = \pa_z \pa_{\bz} K(z,\bz)$.
We define matrix K\"ahler normal coordinate 
$V_{z_0}^z$ at point $z_0$
by simply replacing the ordinary coordinate in (\ref{KNC})
with the matrix coordinate:
\bea
 V^z_{z_0}(Z)
 &=& Z-z_0 + \sum_{N=2}^{\infty} \frac{1}{n!}
 g^{z\bz}(z_0,\bz_0)  \pa_z^n \pa_\bz K(z_0,\bz_0)
   (Z-z_0)^n \nonumber \\
 &=&  \sum_{n=1}^{\infty} \frac{1}{n!}
 g^{z\bz}(z_0,\bz_0)   \pa_z^n \pa_\bz K(z_0,\bz_0)
   (Z-z_0)^n .
 \label{MKNC}
\eea
When $Z$ is diagonal, so is $V^z_{z_0}$ and each
diagonal element is given
by an ordinary K\"ahler normal coordinate.
Under the holomorphic coordinate transformation (\ref{holo}),
the matrix K\"ahler normal coordinate $V^z_{z_0}$ transforms as a
holomorphic tangent vector at point $z_0$:
\bea
 \label{hvec}
V^z_{z_0}(Z) \rightarrow 
V^f_{f_0}(F)= \frac{\pa f}{\pa z}(z_0) V^z_{z_0}(Z)\, .
\eea
The proof is essentially the same to the ordinary case \cite{HIN},
see the appendix.

\subsection{Holomorphically covariant actions}

The powerful result of \cite{ours}
is that once we have a
matrix-valued object which is built from coordinate matrices
and transforms as a tangent vector at each point on
the manifold,
we can construct covariant actions
for multiple D-branes.
We use this result to write down our matrix model actions
in the form of multiple D-instanton actions.

A key object in our construction was 
a matrix distribution
function $\delta(V_{z_0})$ which is defined by
\bea
\label{covdist}
\delta(V_{z_0}) = \int {d \mu d \bmu} \; 
e^{ i(\mu_{\alpha} V^{\alpha}_{z_0}  + \bmu_{\bar{\alpha}} 
\bV^{\bar{\alpha}}_{\bz_0}) } \quad .
\eea
Here,
$d\mu d\bmu \equiv \frac{d\mu_1 d \mu_2}{(2\pi)^2}$
where
$\mu = (\mu_1 + i \mu_2)/2$, $\mu_1, \mu_2 \in {\bf R}$
and
$V^\a_{z_0} = e_z^\alpha (z_0,\bz_0) V^z_{z_0}$ where 
$e_z^\alpha (z_0,\bz_0)$ being zweibein and 
$\alpha$ and $\bar{\alpha}$ are tangent space indices.
$V^\a_{z_0}$ and $\mu_\a$ are scalars under 
holomorphic coordinate transformations
and (co- and contra-variant)
vectors under local Lorentz
transformations.
Thus $\delta(V_{z_0})$ is a 
scalar under holomorphic coordinate
transformations.

In addition to the invariance under 
the holomorphic matrix coordinate
transformation (\ref{holom}), 
we also require invariance under the
$U(N)$ transformation (\ref{UN})
which is supposed to be 
a fundamental symmetry of matrix models.
This is insured if the actions consist of traces of matrices. 
In the following we will only consider single-trace actions,
but the extension to multi-trace case will be similar.

Using the matrix distribution function (\ref{covdist}),
a holomorphically covariant matrix model can be constructed 
with the action
\bea
\label{result}
S[Z,\bZ] 
= \int dz_0 d\bz_0 \,
g_{z\bz}(z_0,\bz_0) \; 
{\rm Tr } 
\left( {\cal L}(V_{z_0}^z(Z),\bV_{\bz_0}^\bz(\bZ), 
g_{z\bz}(z_0,\bz_0))\,
\delta(V_{z_0}) \right) 
\eea
where ${\cal L}(V_{z_0},\bV_{\bz_0}, g(z_0,\bz_0))$
can be an arbitrary scalar built from holomorphic and
anti-holomorphic vectors $V_{z_0}^z$ and $\bV_{\bz_0}^\bz$
and tensors made from metric $g_{z\bz}(z_0,\bz_0)$.
This action is manifestly covariant
under holomorphic coordinate transformations.
Integration over the expansion point $z_0$ here
is natural since configurations of D-instantons
can form higher dimensional branes
which should couple to the metric over 
some region of the space.
However, one can expand the action at a single point
and perform delta function integration explicitly
to obtain an action on the point.
Then the manifest covariance will be lost
but it is useful when one studies 
around a configuration in which 
all D-instantons coincide.
When $V_{z_0}^z$ is diagonal, $\delta(V_{z_0})$
reduces to a diagonal matrix of delta functions
and the action reduces
to a sum of terms corresponding to the individual
D-instantons.

\subsection{Invariant path integral measure}

To define the quantum model by a path integral,
we need to specify the integration measure.
To keep the holomorphic coordinate transformation
as a symmetry, the measure should also be invariant under
the transformation.
The path integral measure for
a single D-instanton is
$\int dz d\bz g_{z\bz}(z,\bz_0) = \int d\bd K(z,\bz)$.
A natural generalization to $N$ multiple D-instantons may be
\bea
 \label{measure}
\int \prod_{(ab)} 
dZ_{ab} d\bZ_{ba} \,
\mbox{det}_{(cd,ef)}\,
\frac{\pa}{\pa Z_{cd}}\frac{\pa}{\pa \bZ_{ef}}
\mbox{Tr}\, K(Z,\bZ)
\eea
where $a,b,\cdots, f$ are matrix indices
and the matrix-valued
K\"ahler potential $ K(Z,\bZ)$ is given by
\bea
 \label{matK}
 K(Z,\bZ) 
\equiv
\sum_{n,m=0}^\infty
\frac{1}{n!}\frac{1}{m!}
\pa_z^n \pa_\bz^m K(z_0,\bz_0) (Z-z_0)^n (\bZ-\bz_0)^m \, .
\eea
$K(Z,\bZ)$ transforms as a scalar under 
holomorphic matrix coordinate transformations: 
$K(Z,\bZ) =  K(F,\bF)$.
This is explicitly checked in the appendix.

To be more precise,
the integration measure for complex matrix $Z$ 
can be more conveniently
understood in terms of forms
\bea
dZ_{ab} \wedge
d\bZ_{ba}\wedge  
dZ_{ba} \wedge
d\bZ_{ab}  
&\propto & 
d \mbox{Re} X^1_{ab} \wedge d \mbox{Im} X^1_{ab} 
\wedge d \mbox{Re} X^2_{ab}\wedge d \mbox{Im} X^2_{ab}
\quad (a>b) 
\nonumber \\
dZ_{aa} \wedge d\bZ_{aa}   
&\propto & 
d X^1_{aa} \wedge d X^2_{aa}
\eea
where $Z=X^1+ i X^2$, and $X^1$,$X^2$ are $N\times N$
hermitian matrices.
The overall numerical factor of the measure
can be chosen appropriately.

The measure (\ref{measure}) is manifestly invariant
under holomorphic matrix coordinate transformations
(\ref{holom}) and $U(N)$ transformations (\ref{UN}).
It is also invariant under the shift of K\"ahler
potential by holomorphic and anti-holomorphic functions:
$K(z,\bz)\rightarrow K(z,\bz)+\psi(z)+\bar{\psi}(\bz)$.
It also reduces to the usual matrix model measure
when ${\cal M}$ is flat $\mbox{\bf R}^2$
with Cartesian coordinates.
The measure (\ref{measure}) is physically natural,
for suppose we have an action which 
makes the path integral strongly localized to a saddle point
given by $[Z,\bZ]=0$.
After the saddle point approximation for off-diagonal elements,
the measure will reduce to
$
\prod_a
dZ_{aa} d\bZ_{aa}\,
g_{z\bz}(Z_{aa},\bZ_{aa}) 
$
which is a suitable measure for
$N$ non-interacting instantons on ${\cal M}$.

Finally, 
using a covariant action constructed by (\ref{result}) 
and the invariant measure (\ref{measure}),
we define a holomorphically covariant
quantum matrix model by the path integral
\bea
 \label{quantum}
\int_{spec\, Z \subset {\cal M}} 
\prod_{(ab)} 
dZ_{ab} d\bZ_{ba} \, 
\mbox{det}_{(cd,ef)}
\left(
\frac{\pa}{\pa Z_{cd}}\frac{\pa}{\pa \bZ_{ef}}
\mbox{Tr}\, K(Z,\bZ)
\right) \,
e^{-S[Z,\bZ]} \, 
\eea
where the path integral 
is performed
over all configurations
satisfying
$spec\, Z \subset {\cal M}$.

\section{Summary and future directions}

In this article we presented a method to
construct quantum
matrix models on arbitrary 
simply connected oriented
real two dimensional
Riemannian manifolds.  
The 
actions 
are manifestly covariant
and proposed path
integral measure is
invariant under
holomorphic matrix coordinate transformations.

Apparently
there are many interesting directions
one can pursue using these models.
One natural direction is 
to study the physics of
non-commuting matrix coordinate configurations
\cite{Sah,gmyers,Hyaku,hasshan}
on non-trivial gravitational backgrounds.
Now that we have concrete models on
general curved backgrounds,
it would be also possible
to study the change of background
under large $N$ renormalization 
\cite{douglas2,BZ}.

In this article we restricted ourselves to 
simply connected manifolds
covered with a single coordinate.
In the case of manifold with
non-trivial topology,
we may need to understand
how to describe the universal covering
in terms of matrix coordinates.
We hope to come back to this issue
in the near future.
There are few simple examples
where the universal coverings
by matrix coordinates are realized \cite{Wati}.

The extension to the higher dimensional case
is a challenging issue.
But it is encouraging that in our one complex dimensional
models we could not only obtain consistent transformation law
and covariant actions but also find the invariant
path integral measure and thus define the quantum models.
For K\"ahler manifolds, it may be the case that
the invariant measure in higher dimensions
takes 
a similar form to that in Eq.~(\ref{measure}).

Although the motivation for studying 
matrix coordinate transformations
came from string theory, we have not used
string worldsheet techniques in this article.
One obstacle is that it is hard to 
solve string theory and classify D-branes
in general gravitational backgrounds.
However, it would be interesting to
study coordinate transformations
in string theory with some particular target space
and D-branes on it.
Interesting clues in this direction can be found
in \cite{Ish2,OO,Target,hassan}.
In particular, Ref.~\cite{Target}
studies non-trivial deformation of
target space diffeomorphism symmetry
in a topological open string theory
on space-filling D-brane 
on Poisson manifolds.
This would be more closely related
to our model if one studies
lower dimensional D-branes in that theory.

What we have done here is a construction 
of some kind of non-commutative
geometry \cite{Connes}.
In particular, there is a close resemblance 
to deformation quantization \cite{Kon},
and string worldsheet calculations cited above
were mostly done in backgrounds which
give rise to deformation quantizations.
On the other hand, there seems to be
some differences between two.
Here, associativity of the product
is manifest, and there's no reference
to a symplectic or Poisson structure.
In deformation quantization
the associative product is the thing to construct
and this refers to the symplectic or Poisson structure.
However, it is known that
the deformation quantization of ${\bf R}^{2d}$
appears in matrix models,
and here although the matrix model actions
do not explicitly contain 
the symplectic structure, it appears 
as a classical solution.
It would be interesting 
to make contact with theories in mathematics.

\section*{Acknowledgments}

We are grateful to Dominic Brecher, 
Henry Ling and Mark Van Raamsdonk for 
the fruitful collaboration \cite{ours},
reading the manuscript and helpful comments.
We would also like to thank the IHES for hospitality 
where this work was completed.
We'd like to thank
Thibault Damour,
Stefan Fredenhagen, 
Ofer Gabber,
Anton Gerasimov, 
Pietro Grassi,
Gyula Karolyi,
Xiabo Liu,
Nikita Nekrasov,
Eric Novac,
Emma Previato,
Sanjaye Ramgoolam, 
Jeong-Hyuck Park,
Zoran Scoda,
Yuji Tachikawa and
Satoshi Yamaguchi
for useful discussions and
stimulating conversations.

\section*{Appendix}
\appendix

To show the transformation laws of 
the holomorphic matrix vector $V_{z_0}^z$
and the matrix-valued K\"ahler potential
$K(Z,\bZ)$ under 
holomorphic matrix coordinate transformations,
we first prove the following formula.
It is presented in a form
applicable to higher dimensional case.
In the following, indices after 
a comma denote partial differentiations.
For simplicity,
let us consider a holomorphic coordinate transformation
$z^i \to z'{}^i = z'{}^i(z)$ which keeps the origin.
Extension to the case that doesn't keep the origin is obvious.
We shall show that
the transformation law of
$K,_{i_1 \cdots i_{n'} \bi_1 \cdots \bi_{m'}} (z,\bz)$
($n',m' \geq 1$)
under the holomorphic coordinate transformation 
is given by
\bea
 &&K,_{i_1 \cdots i_{n'}  \bi_1 \cdots \bi_{m'}  }(z,\bz) \nonumber \\
 &\rightarrow &
 K,_{i_1' \cdots i_{n'}'  \bi_1' \cdots \bi_{m'}'}(z',\bz') \nonumber \\
&&= \sum_{n=1}^{n'} \sum_{m=1}^{m'} \frac{1}{n !} \frac{1}{m !} 
K,_{k_1 \cdots k_n \bk_1 \cdots \bk_m} (z,\bz)
 \left[ {\pa^{n'} (z^{k_1} \cdots z^{k_n})
  \over \pa z'{}^{i_1'} \cdots \pa z'{}^{i_{n'}'} } \right]_* 
 \left[ {\pa^{m'} (\bz^{\bk_1} \cdots \bz^{\bk_m})
  \over \pa \bz'{}^{\bi_1'} \cdots \pa \bz'{}^{\bi_{m'}'} } \right]_* 
 \label{lemma}
\eea
where $[\cdots]_*$ means that
terms including $z$ ($\bz$) that are differentiated by no 
$z'$ ($\bz'$)
are omitted.
The proof is essentially the same to the one in \cite{HIN}
and presented here for readers' convenience. 
We show Eq.~(\ref{lemma}) by induction.\\
i) In the case ${n'}={m'}=1$, Eq.~(\ref{lemma}) means
\bea
 \label{nm1}
K,_{i'\bi'} (z',\bz')=
K,_{i\bi} (z,\bz)
\frac{\pa z^i}{\pa z'{}^{i'}}
\frac{\pa \bz^{\bi}}{\pa \bz'{}^{{\bi}'}}
\eea
which follows from the chain rule for differentiation.
(\ref{nm1}) is the transformation law for
the K\"ahler metric. \\
ii) We assume that Eq.~(\ref{lemma}) holds for ${n'},{m'}$.
Differentiation of Eq.~(\ref{lemma})
by $z'{}^{{i'}_{{n'}+1}}$ gives
\bea
&& K,_{i_1'\cdots i_{{n'}+1}'\bi_1' \cdots \bi_{m'}' } \nonumber \\
&=&
 \sum_{n=1}^{n'} \sum_{m=1}^{m'}
 \frac{1}{n !} \frac{1}{m !} 
 K,_{k_1 \cdots k_{n+1}\bk_1 \cdots \bk_m }
 {\pa z^{k_{n+1}} \over \pa z'{}^{i_{{n'}+1}'}}
 \left[ {\pa^{n'} (z^{k_1} \cdots z^{k_n})
  \over \pa z'{}^{i_1'} \cdots \pa z'{}^{i_{n'}'} } \right]_* 
 \left[ {\pa^{m'} (\bz^{\bk_1} \cdots \bz^{\bk_m})
  \over \pa \bz'{}^{i_1'} \cdots \pa \bz'{}^{i_{m'}'} } \right]_* 
\nonumber \\
&& + \sum_{n=1}^{n'} \sum_{m=1}^{m'}
 \frac{1}{n !} \frac{1}{m !} 
  K,_{k_1 \cdots k_n \bk_1 \cdots \bk_m }
 {\pa \over \pa z'{}^{i_{{n'}+1}'}}
 \left[ {\pa^{n'} (z^{k_1} \cdots z^{k_n})
  \over \pa z'{}^{i_1'} \cdots \pa z'{}^{i_{n'}'} } \right]_*
\left[ {\pa^{m'} (\bz^{\bk_1} \cdots \bz^{\bk_m})
  \over \pa \bz'{}^{\bi_1'} \cdots \pa \bz'{}^{\bi_{m'}'} } \right]_* 
.\nonumber \\
\eea
The first term can be rewritten as
\bea
 \sum_{n=2}^{{n'}+1} \sum_{m=1}^{m'} {1 \over (n-1)!}{1 \over m!}
 K,_{k_1 \cdots k_n \bk_1 \cdots \bk_m }
 {\pa z^{k_n} \over \pa z'{}^{i_{{n'}+1}'}}
 \left[ {\pa^{n'} (z^{k_1} \cdots z^{k_{n-1}})
  \over \pa z'{}^{i_1'} \cdots \pa z'{}^{i_{n'}'} } \right]_* 
\left[ {\pa^{m'} (\bz^{\bk_1} \cdots \bz^{\bk_m})
  \over \pa \bz'{}^{\bi_1'} \cdots \pa \bz'{}^{\bi_{m'}'} } \right]_* 
\,.
\eea
Therefore, we have
\bea
&& K,_{i_1'\cdots i_{{n'}+1}' \bi_1'\cdots \bi_{{m'}}'}\nonumber \\
&=& 
\sum_{m=1}^{m'} {1 \over m!}
 K,_{k_1 \bk_1 \cdots \bk_m}
 {\pa^{{n'}+1} z^{k_1}
  \over \pa z'{}^{i_1'} \cdots \pa z'{}^{i_{{n'}+1}'} } 
 \left[ {\pa^{m'} (\bz^{\bk_1} \cdots \bz^{\bk_m})
  \over \pa \bz'{}^{\bi_1'} \cdots \pa \bz'{}^{\bi_{m'}'} } \right]_* 
\nonumber \\
 &&+ \sum_{n=2}^{n'}
\sum_{m=1}^{m'} {1 \over n!}{1 \over m!}
 K,_{k_1 \cdots k_n \bk_1 \cdots \bk_m}
 \left\{
 n {\pa z^{k_n} \over \pa z'{}^{i_{{n'}+1}}}
 \left[ {\pa^{n'} (z^{k_1} \cdots z^{k_{n-1}})
  \over \pa z'{}^{i_1'} \cdots \pa z'{}^{i_{n'}'} } \right]_*
 \left[ {\pa^{m'} (\bz^{\bk_1} \cdots \bz^{\bk_m})
  \over \pa \bz'{}^{\bi_1'} \cdots \pa \bz'{}^{\bi_{m'}'} } \right]_* 
 \right. \nonumber \\
 &&
 \qquad \qquad \qquad \qquad \qquad \qquad 
  \left.
 + {\pa \over \pa z'{}^{i_{{n'}+1}'}}
 \left[ {\pa^{n'} (z^{k_1} \cdots z^{k_n})
  \over \pa z'{}^{i_1'} \cdots \pa z'{}^{i_{n'}'} } \right]_*
 \left[ {\pa^{m'} (\bz^{\bk_1} \cdots \bz^{\bk_m})
  \over \pa \bz'{}^{\bi_1'} \cdots \pa \bz'{}^{\bi_{m'}'} } \right]_* 
 \right\} \nonumber \\
&&+ 
\sum_{m=1}^{m'}
\frac{1}{{n'} !} {1 \over m!}
 K,_{k_1 \cdots k_{{n'}+1} \bk_1 \cdots \bk_m}
 {\pa z^{k_{{n'}+1}} \over \pa z'{}^{i_{{n'}+1}'}}
 \left[ {\pa^{n'} (z^{k_1} \cdots z^{k_{n'}})
  \over \pa z'{}^{i_1'} \cdots \pa z'{}^{i_{n'}'} } \right]_* 
 \left[ {\pa^{m'} (\bz^{\bk_1} \cdots \bz^{\bk_m})
  \over \pa \bz'{}^{\bi_1'} \cdots \pa \bz'{}^{\bi_{m'}'} } \right]_*
\,.
\eea
For the terms in the curly brackets,
we apply the relation
\bea
 && n {\pa z^{k_n} \over \pa z'{}^{i_{{n'}+1}'}}
 \left[ {\pa^{n'} (z^{k_1} \cdots z^{k_{n-1}})
  \over \pa z'{}^{i_1'} \cdots \pa z'{}^{i_{n'}'} } \right]_*
 + {\pa \over \pa z'{}^{i_{{n'}+1}'}}
 \left[ {\pa^{n'} (z^{k_1} \cdots z^{k_n})
  \over \pa z'{}^{i_1'} \cdots \pa z'{}^{i_{n'}'} } \right]_* \nonumber \\
 && = \left[ {\pa^{{n'}+1} (z^{k_1} \cdots z^{k_n})
  \over \pa z'{}^{i_1'} \cdots \pa z'{}^{i_{{n'}+1}'} } \right]_* \
\eea
where symmetrization of the first term
on the left-hand side is implied.
Then,
\bea
&& K,_{i_1'\cdots i_{{n'}+1}'\bi_1'\cdots \bi_{{m'}}'}
\nonumber \\
&& = \sum_{n=1}^{{n'}+1} \sum_{m=1}^{{m'}} 
 \frac{1}{n !} \frac{1}{m !}
 K,_{k_1 \cdots k_n \bk_1 \cdots \bk_m}
 \left[ {\pa^{{n'}+1} (z^{k_1} \cdots z^{k_n})
  \over \pa z'{}^{i_1} \cdots \pa z'{}^{i_{{n'}+1}} } \right]_* 
 \left[ {\pa^{m'} (\bz^{\bk_1} \cdots \bz^{\bk_m})
  \over \pa \bz'{}^{\bi_1'} \cdots \pa \bz'{}^{\bi_{m'}'} } \right]_*.
\eea
Thus Eq.~(\ref{lemma}) holds for ${n'}+1,{m'}$.
That Eq.~(\ref{lemma}) also holds for ${n'},{m'}+1$ 
can be shown similarly.
From i) and ii) Eq.~(\ref{lemma}) is proved.

Now let us go back to our one complex dimensional case.
From Eq.~(\ref{lemma}) we can immediately see that
under holomorphic coordinate transformations,
$\pa_z^{n'} \pa_\bz^{m'} K$ transforms according to
\bea
 \label{TKT}
&& \pa_z^{n'} \pa_\bz^{m'} K|_{0} \nonumber \\
&\rightarrow &
 \pa_f^{n'} \pa_\fb^{m'} K|_{0}
 = \sum_{n=1}^{n'} \sum_{m=1}^{m'} 
\frac{1}{n !}\frac{1}{m!} \pa_z^n \pa_\bz^m K|_{0}
 \left[
 {\pa^{n'} (z^n)
  \over \pa^{n'} f} \right]_{0}
 \left[
 {\pa^{m'} (\bz^m)
  \over \pa^{m'} \fb} \right]_{0}  \nonumber \\
 &&\qquad \qquad \,\,\,\, = \sum_{n=1}^{\infty}  \sum_{m=1}^\infty
 \frac{1}{n !}\frac{1}{m !} \pa_z^{n}\pa_\bz^{m} K|_{0}
 \left[
 {\pa^{n'} (z^n)
  \over \pa^{n'} f} \right]_{0}
\left[
 {\pa^{m'} (\bz^m)
  \over \pa^{m'} \fb} \right]_{0} . 
 \label{corollary}
\eea
Here, the subscript $0$ denotes
that it is a value at the origin.
The second equality holds because
the terms in $[\cdots]_{0}$ 
vanish when $n>{n'}, m>{m'}$.


Using
Eq.~(\ref{corollary})
(the third line for the holomorphic part and
the second line for the anti-holomorphic part),
the left-hand side of 
Eq.~(\ref{hvec}) can be explicitly
calculated from the definition (\ref{KNC}):
\bea
 \label{Ahvec}
 V^f_0(F)
&=& \sum_{n=1}^{\infty} \frac{1}{n!}
 g^{f\fb} \pa_{f}^n \pa_{\fb} K|_0
  F^n \nonumber \\
&=& \sum_{n=1}^{\infty} \frac{1}{n!} g^{f\fb}|_0
 \left(\sum_{m=1}^{\infty} \frac{1}{m !} 
  \pa_z^m \pa_\bz K|_0
 \left[
 {\pa \bz \over \pa \fb}
 \right]_0
 \left[
 {\pa^n (z^m)
  \over \pa^n f } \right]_{0}\right)
 F^n \nonumber \\
&=& {\pa f \over \pa z}(0)
 \sum_{n=1}^{\infty} \sum_{m=1}^{\infty} 
 \frac{1}{n!}
 \frac{1}{m!}
 g^{z\bz}  \pa_z^m \pa_\bz K|_0
 \left[{\pa^n (z^m)
  \over \pa^n f}  \right]_0
   F^n \nonumber \\
&=& {\pa f \over \pa z}(0)
 \sum_{m=1}^{\infty} \frac{1}{m!}
 g^{z\bz} \pa_z^m \pa_\bz K|_0
 Z^m
= {\pa f \over \pa z}(0)  V^z_0(Z) \,.
\eea
Thus, under holomorphic coordinate transformations
the matrix K\"ahler normal coordinate
$V^z_{z_0}$ transforms as a holomorphic tangent vector
at point $z_0$.

It is also straightforward to show that
the matrix-valued K\"ahler potential
$K(Z,\bZ)$ transforms as a scalar:
\bea
 \label{scalarK}
K(F,\bF) 
&=&
\sum_{{n'},{m'}=0}^\infty
\frac{1}{{n'}!}\frac{1}{{m'}!}
\pa_f^{n'} \pa_\fb^{m'} K|_0 F^{n'} \bF^{m'} 
\nonumber \\
&=&
K_0 +
\pa_z K|_0 \frac{\pa z}{\pa f}(0) F +
\pa_\bz K|_0 \frac{\pa \bz}{\pa \fb}(0) \bF 
\nonumber \\
&&+
\sum_{{n'},{m'}=1}^\infty
\frac{1}{{n'}!}\frac{1}{{m'}!}
\sum_{n,m=1}^\infty
\frac{1}{n!}\frac{1}{m!}
 \pa_z^{{n}}\pa_\bz^{{m}} K|_{0}
 \left[
 {\pa^{{n'}} (z^n)
  \over \pa^{{n'}} f} \right]_{0}
\left[
 {\pa^{{m'}} (\bz^m)
  \over \pa^{{m'}} \fb} \right]_{0} 
F^{n'} \bF^{m'}
\nonumber \\
&=&
\sum_{n,m=0}^\infty
\frac{1}{n!}\frac{1}{m!}
 \pa_z^{{n'}}\pa_\bz^{{m'}} K|_{0}
Z^n \bZ^m 
\nonumber \\
&=& K(Z,\bZ) \, .
\eea


\end{document}